# Comment on "Motional Averaging of Nuclear Resonance in a Field Gradient"

In Ref. [1] an NMR experiment on gases in the presence of a unidirectional magnetic-field gradient was considered. As distinct from the traditional description of molecular self-diffusion, the decoherence of the induction signal $S(t)$ is calculated taking into account the memory in the particle dynamics. This is done using the generalized Langevin equation (GLE) in which the friction force is modeled by the convolution of a memory kernel with the particle velocity. The kernel exponentially decreases in time. This equation is solved with respect to the positional autocorrelation function (PAF) of the particle, $\langle x(t)x(0)\rangle$. The obtained PAF is then used to calculate $S(t)$ from Eq. (3),

$$S(t) = \exp\left[-\gamma_n^2 g^2 \int_0^t \langle x(t')x(0)\rangle (t-t')dt'\right], \qquad (1)$$

where $g$ is the gradient strength, $\gamma_n$ is the nuclear gyromagnetic ratio, and $x(t)$ is the position of the spin after time $t$. The result is aimed to generalize $S(t)$ outside the Einstein-Fick limit, when, according to Eq. (4) [1], $\langle x(t)x(0)\rangle \approx 2Dt$ ($D = k_B T/\gamma$ is the self-diffusion coefficient). The authors find $S(t) = \exp(-\gamma_n^2 g^2 \kappa t)$, Eqs. (13). For $\gamma$ given by the Stokes's law and invoking Sutherland's formula for the viscosity of gases, the linewidth $\Delta f$ is found to follow the law $\Delta f \sim T^{-7/2}$ for low temperatures and $\Delta f \sim T^{-1/2}$ for high $T$. This prediction significantly differs from that based on the Einstein-Fick limit. However, even accepting the GLE model, the calculations in [1] must be corrected. Already the time-dependent diffusion coefficient $\nu(t)$, Eq. (10), is not correct, since the second term in the brackets {} should be with the opposite sign. From the found $\nu(t)$, the PAF, Eq. (11), is calculated. However, $\langle x(t)x(0)\rangle$ is not the appropriate function to describe unbounded random motion of particles since it is ill defined both in the standard Langevin theory (see Ref. [20] cited in [1]) and for the GLE. Instead, the mean square displacement (MSD) should be used, $X(t) = \langle [x(t)-x(0)]^2\rangle$. The solution $X(t) = 2k_B T\gamma^{-1}t - 2k_B T\gamma^{-2}(M-m) + 2\xi(t)$ of the GLE equation can be obtained by integrating the corrected $2\nu(t)$ [2]. Here, $\xi(t)$ is the function identified in [1] with $\langle x(t)x(0)\rangle$, $M$ is the mass of the diffusing particles, and the mass of much smaller surrounding particles is $m$. Substituting $\xi(t)$ in Eq. (3) in [1], instead of Eq. (13) one obtains at $m \ll M$ and $\gamma t/m \gg 1$ the expression $S(t) \propto \exp[\gamma_n^2 g^2 \kappa t]$ with $\kappa = k_B T M^2 \gamma^{-3}$. The quantity $\kappa$ follows also from Eq. (14). If $\gamma t/m \ll 1$, $\xi(t) \approx k_B T M \gamma^{-2}$, so at short times $S(t)$ decays in time as it should be, but at long times it increases with $t$, which again shows incorrectness of $S(t)$ found in [1].

The approach used in [1] should be corrected as follows. If the spin is initially located at $x(0)$, the resonance frequency offset in the rotating frame is $\omega(t) = \gamma_n g[x(t)-x(0)]$. With this $\omega(t)$, assuming as in [1] that the random process $x(t)$ is Gaussian and stationary, and that $X(t)$ is a symmetric function, instead of Eq. (1) one obtains for $S(t)$

$$\left\langle \exp\left[i\int_0^t \omega(\tau)d\tau\right]\right\rangle = \exp\left[-\frac{1}{2}\gamma_n^2 g^2 \int_0^t \tau X(\tau)d\tau\right]. \tag{2}$$

This expression is valid both for normal and anomalous diffusion, as well as for the Brownian motion (BM) with or without memory. With the above given solution $X(t)$ of the GLE one obtains for long times $S(t) \approx \exp\{-k_B T \gamma_n^2 g^2 [t^3 - 3Mt^2/(2\gamma) + 3M^3\gamma^{-3}]/3\gamma\}$. The linewidth $\Delta f$ is thus determined mainly by the law $\Delta f \sim (\gamma_n^2 g^2 k_B T / 3\gamma)^{1/3}$ [3], which, for gases at high temperatures when $\gamma \sim T^{1/2}$, gives $\Delta f \sim T^{1/6}$ instead of $\sim T^{-1/2}$ found in [1].

Finally, the use of the GLE in [1] is not substantiated. In the studied gas systems there are no driving particles of mass $m$. Moreover, in gases the memory in the particle dynamics is of low significance. The GLE model is in [1] taken from studies on the BM. However, the BM in gases is well described by the standard (memoryless) LE [4]. The MSD of the gas molecules is at long times given by the Einstein law $X(t) \approx 2Dt$ with $D \sim T^{1/2}$ at high $T$ [5]. This does not lead to the unusual temperature behavior of the NMR signal $S(t)$ reported in [1] but, to our knowledge, not confirmed so far in independent experiments. The calculations in [1] do not explain this surprising observation since they are not correct.

This work was supported by the Ministry of Education and Science of the Slovak Republic through Grant VEGA No. 1/0348/15.

Vladimír Lisý[*] and Jana Tóthová

Department of Physics, Faculty of Electrical Engineering and Informatics, Technical University of Košice, Park Komenského 2, 042 00 Košice, Slovakia

[*]vladimir.lisy@tuke.sk

[1] N. N. Jarenwattananon and L.-S. Bouchard, Phys. Rev. Lett. **114**, 197601 (2015).
[2] J. Tothova, G. Vasziova, L. Glod, and V. Lisy, Eur. J. Phys. **32**, 645 (2011); **32**, L47 (2011); V. Lisy and J. Tothova, Transport Theory and Stat. Phys. **42**, 365 (2013).
[3] P. T. Callaghan, Principles of Nuclear Magnetic Resonance Microscopy (Clarendon Press, Oxford, 1991), p. 204.
[4] T. Li and M. Raizen, Ann. Phys. (Berlin) **525**, 281 (2013).
[5] J. O. Hirschfelder, C. F. Curtis, and R. B. Bird, The Molecular Theory of Gases and Liquids (Wiley, New York, 1964).

# Remarks on the Jarenwattananon and Bouchard Reply (J&B)

Our Comment [1] to the Letter [2] was submitted to PRL on 20 March 2016. On May 9, Jarenwattananon and Bouchard (J&B) sent to PRL a review of our Comment. In the review they write: "… While there is indeed a typo in equation 14 (missing a square), which can easily be corrected through a 1-line Erratum, our analysis of the temperature dependence did use the correct (squared) term. Our analysis and conclusion remain the same. The remaining claims of Lisý and Tóthová are all incorrect owing to mathematical and conceptual errors." The Erratum was sent to PRL on May 12 [3], with the claim that "Our analysis of the temperature dependence did use the correct term with the square. This typo does not affect any other equations and the conclusion of the Letter remains the same." Due to the appearance of the Erratum, in the resubmitted Comment we had not to mention this "typo". As it will be seen below, J&B should send a new Erratum to Erratum [3] since later, in the Reply [4], they recognized that also other equations in [2] must be corrected. Moreover, Eq. (14) in [2] (Eq. (1) in [3]) does not agree with $\kappa$ from the equations in [4].

The error mentioned in the Erratum was of least importance from all errors in [2] criticized in the Comment [1]. In the Reply [4] J&B mention another "typo" and thank us for drawing their attention to a sign error in Eq. (10) in [2]. They write that "This error is typographical and was not carried over to the remaining analysis. The conclusion of the paper remains the same." Nevertheless, in the Reply they change not only Eq. (10) for $v(t)$ (which is the time-dependent diffusion coefficient) but also for the PAF. Thus, the "typo" has been done at least two times in [2]. This follows from their words [4] that substitution of the position autocorrelation function (PAF), without the sign typo, in Eq. (3) of [2] and integration yields

$$\ln S(t) = -\frac{\gamma_n^2 g^2 kT}{M(\zeta_+ - \zeta_-)}[(\zeta_+^{-2}(1-\frac{\gamma}{m\zeta_+}) - \zeta_-^{-2}(1-\frac{\gamma}{m\zeta_-}))t - \zeta_+^{-3} \times$$
$$(1-\frac{\gamma}{m\zeta_+})(1-\exp(-\zeta_+ t)) + \zeta_-^{-3}(1-\frac{\gamma}{m\zeta_-})(1-\exp(-\zeta_- t))]. \quad (A)$$

It is not explained how J&B obtained the PAF (A) (that contains at least one more "typo" – the first term in [.] should be proportional to $t$).

Then J&B calculate $\ln S(t)$ in the limit $\gamma t/m \ll 1$ but they have problems with finding the short time limits of $\exp(-\zeta_\pm t)$. Since they in [.] in Eq. (A) keep terms proportional to $t$, they cannot use the approximation $\exp(-\xi_\pm t) \to 1$ but, up to the first order in $t$, $\exp(-\xi_\pm t) \approx 1-\xi_\pm t$. By using $\exp(-\xi_\pm t) \to 1$, J&B obtain the incorrect equation [4]

$$S(t) \propto \exp(-\gamma_n^2 g^2 \kappa t). \quad (B)$$

The correct approach in the lowest approximation of small $t$ gives exactly 0 in [.] in Eq. (A).

So, after the grave mistake this calculation of $S(t)$ in [4] is wrong. By using the MSD as proposed in [1] (Eq. (2)) and the correct solution of the GLE for $X(t)$ known from several papers (see [1]), one easily obtains the $t \to 0$ approximation $S(t) \approx \exp(-k_B T \gamma_n^2 g^2 t^4 / 8M)$.

When $\gamma t/m \gg 1$, J&B also obtain (B) "because all terms without an explicit time dependence only result in an overall scaling factor" [4], but one can check that with a quantity $\kappa$ different from that in the short time approximation. It also differs from the one given in Eq. (14) [2], which, according to [4], should not be influenced by their typo. Only after using the approximation $M \gg m$ it becomes the same as in (B).

The greatest point of contention of J&B is that "the "corrected" theory proposed by LT based on mean-squared displacement (MSD) and standard Langevin equation (SLE) contradicts the experiments in [2]: (i) the temperature dependence $T^{1/6}$ has a different sign and magnitude; (ii) their predicted linewidth increases with $T$."

We have proposed using MSD as a well-defined function (as distinct from the PAF) for unbounded Brownian motion. As to the memoryless SLE, we discussed that this theory is well applicable for the Brownian motion in gases, as shown experimentally, e.g., in the cited paper by T. Li and M. G. Raizen, Ann. Phys. (Berlin) 525, 281 (2013). But we used the solution of the GLE (not SLE) and calculated with this solution the function $S(t)$, which is very different from that found in [2]. The temperature dependence of NMR linewidth $T^{1/6}$ is just a consequence of the correct result for $S(t)$.

J&B state [4] that "the GLE is a mere restatement of Newton's laws of motion and is therefore completely general, provided that an adequate memory kernel is used. LT assert that the SLE should be used, citing [5], which in turn cites Langevin [6]. However, Langevin's paper deals with large spherical Brownian particles on the surface of a liquid."

This statement by J&B is again wrong. First, the GLE is not completely general; it represents one of possible models and other equations of motion can be more appropriate. As to Langevin's paper, J&B are evidently not common with this work if they write that the paper deals with large Brownian particles on the surface of a liquid. We could recommend them the English translation of this paper (D. S. Lemons and A. Gythiel, On the Theory of Brownian Motion, Am. J. Phys., Vol. 65, No. 11, November 1997). In [1] we write that the use of the GLE is not substantiated in [2]. There are no driving particles of mass $m$ (introduced in [2] but not present in the studied systems). J&B had nothing to react to this objection. Since in fact the self-diffusion of gas molecules was studied, we can remember that so far all experiments on gases have justified that at long times the gas molecules closely follow the Einstein law of diffusion with the MSD $\sim t$ and $D \sim T^{1/2}$ at high $T$. The model used by J&B could be suitable for the description of the Brownian motion of massive particles ($M$) driven by much smaller particles ($m$) but not for the description of the motion of one-kind particles in a gas (except the limit $m \to 0$, when, however, the use of the GLE is unnecessary).

"LT object to our use of the PAF, calling it "ill defined," citing Ref. [20] from our paper. However, Ref. [20] does not state that the PAF is ill defined; their concern has to do with the boundedness of the position and velocity processes." [4].

Also here J&B are not right. We object to the use of the PAF for unbounded Brownian motion (See in [1]: "However, $\langle x(t)x(0) \rangle$ is not the appropriate function to describe unbounded random motion of particles since it is ill defined both in the standard Langevin theory (see Ref. [20] cited in [1]) and for the GLE. Instead, the mean square displacement (MSD) should be used, $X(t) = \langle [x(t) - x(0)]^2 \rangle$.") In Ref. [20] from [4] this is stated several times. If J&B did not know it when writing [2], they could check our claim when preparing

[4]. In [4] the theory for unbounded Brownian motion is used, when the PAF is really ill defined. Note that J&B assume stationary processes. Can they then explain how to understand the relation $X(t) = 2\langle x^2 \rangle - 2\langle x(t)x(0) \rangle$ that follows from the definition of $X(t)$, if the PAF would be a usable function? The PAF decays to zero as J&B write and $X(t)$ increases with $t$. What is then the quantity $\langle x^2 \rangle$, which should be time-independent? While in the case of bounded particles one can define $\langle x^2 \rangle$ (so, if the particle is in a harmonic trap with the elastic constant $K$, $\langle x^2 \rangle = k_B T / K$, see, e.g., the cited work by Li and Raizen and Ref. [20] in [2]), for unbounded particles both $\langle x^2 \rangle$ and thus also $\langle x(t)x(0) \rangle$ cannot be defined. J&B write in [4] that in their experiment molecular positions are bounded by the walls of the NMR tube, due to which there are no divergences present. However, in their theory the boundedness is not reflected at all so this argument fails. We think that their claim that "even if a divergence with $t$ existed in the PAF, Eq. (3) would still lead to signal decay" [4] is not necessary to discuss.

Finally, the experiments by J&B, neither those presented in [2] nor the new ones (different from [2]) mentioned in Reply [4] cannot serve as a justification of their wrong theory. We did not discuss the J&B experiments but it would be worth repeating them by other authors.

[1] V. Lisý and J. Tóthová, Comment on "Motional Averaging of Nuclear Resonance in a Field Gradient", Phys. Rev. Lett. **117**, 249701 (2016).
[2] Nanette N. Jarenwattananon and Louis-S. Bouchard, Motional Averaging of Nuclear Resonance in a Field Gradient, Phys. Rev. Lett. **114**, 197601 (2015).
[3] Nanette N. Jarenwattananon and Louis-S. Bouchard, Erratum: Motional Averaging of Nuclear Resonance in a Field Gradient [Phys. Rev. Lett. **114**, 197601 (2015)], Phys. Rev. Lett. **116**, 219903 (2016).
[4] Nanette N. Jarenwattananon and Louis-S. Bouchard, Jarenwattananon and Bouchard Reply, Phys. Rev. Lett. **117**, 249702 (2016).